# Sharp and Bright Photoluminescence Emission of Single Crystalline Diacetylene Nanoparticles

Seokho Kim[#,a], Piao Xianling[#,b], Hyeong Tae Kim[a], Chunzhi Cui[c],

and Dong Hyuk Park[*,a],

[a]*Department of Applied Organic Materials Engineering, Inha University, Incheon 402-751, Korea.*

[b]*Affiliated Hospital of Yanbian Univertisy, Yanji 133000, China.*

[c]*Department of Chemistry, Yanbian University, Yanji 133002, China.*

Amorphous nanoparticles (NPs) of diacetylene (DA) molecules were prepared by using a reprecipitation method. After crystallization through solvent-vapor annealing process, the highly crystalline DA NPs show different structural and optical characteristics compared with the amorphous DA NPs. The single crystal structure of DA NPs was confirmed by high-resolution transmission electron microscopy (HR-TEM) and wide angle X-ray scattering (WAXS). The luminescence color and photoluminescence (PL) characteristics of the DA NPs were measured using color charge-coupled device (CCD) images and high-resolution laser confocal microscope (LCM). The crystalline DA NPs emit bright green light emission compared with amorphous DA NPs and the main PL peak of the crystalline DA NPs exhibits relative narrow and blue shift phenomena due to enhanced interaction between DA molecular in the nano-size crystal structure.

Keywords: PCDA, solvent-vapor annealing, crystallinity, photoluminescence.

\* Corresponding author. Tel.: +82 32 860 7496; Fax: +82 32 873 0181.

*E-mail address*: donghyuk@inha.ac.kr (D. H. Park).

# These authors contributed equally to this work.

## I. INTRODUCTION

In recent years, functional organic nanomaterials have attracted increasing research interest, due to their potential use in optoelectronics [1,2], nonlinear optics [3,4], and photonics [5,6]. For example, practically, organic NPs are being used as developing materials such as display element [7], ink toners [8], drugs [9], and so on. As a part of these organic nanomaterials, organic crystals, whose optical and electrical properties are fundamentally different from those of inorganic metals and semiconductors due to weak intermolecular interaction forces of the van der Waals type [10] have captured more and more interest. These noncovalent intermolecular interactions are all able to influence strongly the final packing structure [11-14]. Therefore, understanding and controlling organic molecular arrangements in the solid state are fundamental issues for obtaining the desired chemical and physical properties [15]. Furthermore, the key step in nanomaterial research is how to achieve chemical and physical properties of an individual nanomaterial is also became a hot issue in these years [16,17]. Most of studies have been done by ensemble measurement of the colloidal solution, so the spectra are averaged over different size and shape. Therefore, it has been strongly desired to reveal optical properties for individual nanomaterial under dry condition. As the solution, using nanoscale optical equipment such as near-field scanning optical microscope (NSOM) and laser confocal microscope (LCM) are consider as efficient method [18,19]. In the past decade, several works using single NP spectroscopy have been reported for organic nanoparticles using high resolution microscopy optical setup.

In this study, we used 10, 12-pentacosadiynoic acid (PCDA)-based DA molecules as a standard material to construct organic single crystal NPs. However, most of research about the DA molecules was focused on their polymeric phase [20-22].

The amorphous DA NPs were prepared by a conventional reprecipitation method [23,24].

As a crystallization strategy, solvent-vapor annealing method was applied on a slide glass substrate, transformed the amorphous phase of the DA NPs to the single crystalline phase. Then we observed a highly enhanced brightness, a relative narrow and blue-shifted LCM PL spectra of the crystalline DA NPs compared with the amorphous DA NPs. This phenomenon was originated from enhanced interaction between DA molecules in the nanosize single crystalline structure that consider for an effective exciton transport [17].

## II. EXPERIMENTAL

We prepared DA NPs by conventional reprecipitation method. The DA powders were dissolved in tetrahydrofuran (THF) solution as concentration of 1 mg/ml, and then injected 600 μl into vigorously stirred deionized (DI) water (20 ml) followed by filtration with 0.2 μm PTFE filter. For the purpose of crystallization, the dispersion was dried on a slide cover-glass substrate under vacuum (1 bar) at 180 ºC for 10 h.

The formation of the amorphous and crystalline DA NPs was visualized by HR-TEM (JEOL, JEM-3010). The WAXS patterns were collected using Bruker GADDS diffractometer with Cu Kα radiation (λ=1.5418 Å). The luminescence color charged coupled device (CCD) images of the amorphous and crystalline DA NPs were measured with an AVT Marlin F-033C ($\lambda_{ex}$ = 435 nm) and the exposure time was fixed at 0.1 s. The solid-state PL spectra of two kinds of DA NPs were measured by using a LCM built around an inverted optical microscope (Axiovert 200, Zeiss GmbH) which has an unpolarized Argon ion laser (488 nm). The laser power incident on the sample and the acquisition time for each LCM PL spectra were fixed at 1 μW and 1s, respectively. And the spot size of the focused laser beam on the sample was estimated at 190 nm. More details of LCM PL experiments were reported earlier

[25,26].

## III. RESULTS and DISCUSSTION

The formation of the amorphous and crystalline DA NPs was confirmed through HR-TEM as shown in Fig. 1 (a, b) and (c, d). From the HR-TEM images, we observed homogeneous spherical shape NPs in both of the amorphous and crystalline DA NPs and the mean diameters of the amorphous DA NPs and crystalline DA NPs were estimated to be 40 (±10) nm and 30 (±10) nm, respectively. Periodic patterns of stripes are clearly observed for the crystalline DA NPs in Fig. 1 (d) compared to the amorphous DA NPs in Fig. 1 (b). And the lattice spacing value of the crystalline DA NPs is ~2.75 Å, which is the molecular stacking distance. To confirm and support the crystallinity of the crystalline DA NPs, the WAXS patterns of the amorphous and crystalline DA NPs were collected, as shown in Fig. 2. A strong crystalline peak at 31° which corresponding to (311) reflection was observed for crystalline DA NPs and there were no significant peaks for the amorphous DA NPs, indicating the poor crystalline structure of the amorphous DA NPs. Using the Bragg's law based on the WAXS pattern of the crystalline DA NPs, the lattice constant was estimated to be 2.75 Å, which is in agreement with the result of HR-TEM and indicate that the DA molecules in the crystalline DA NPs have a strong molecular interaction. Furthermore, the HR-TEM image and WAXS patterns suggest that the solvent vaporization through the annealing process transformed the the amorphous DA NPs to the crystalline DA NPs.

The nanoscale luminescence characteristics of the amorphous and crystalline DA NPs were investigated by color CCD and LCM PL spectra. Fig. 3 (a) and (b) show the luminescence color CCD images of the amorphous and crystalline DA NPs respectively. From the images, we can observe that the amorphous DA NPs emitted weak pale-green light, while the crystalline DA NPs emitted much brighter green light.

Corresponding LCM PL spectra of the amorphous and crystalline DA NPs are shown in Fig. 4. The LCM spectra were measured at 10 different points, and then averaged to obtain the final PL spectra, and each point estimated in the same condition. Fig. 4 (a) shows the normalized LCM PL spectra of the amorphous and crystalline DA NPs. The main LCM PL peak for the amorphous DA NPs was detected as a broad shape at around 570 nm. Compared with the amorphous DA NPs, the main LCM PL peak of the crystalline DA NPs was blue-shifted and observed as a sharp shape at around 550 nm. The full width half maximum (FWHM) of the normalized LCM PL spectra of the crystalline DA NPs was narrowed significantly compared with the amorphous DA NPs. The FWHM of the PL spectra of the amorphous DA NPs was estimated to be about 157 nm while that of the crystalline DA NPs was about 51 nm. The blue-shifted for the crystalline DA NPs is due to the nano-size crystalline structure effect, and the narrowed LCM PL spectra of the crystalline DA NPs is originated from the light emission of the structural confinement bands due to the single crystal structure [27-29]. Fig. 4 (b) shows the absolute LCM PL spectra of the amorphous and crystalline DA NPs, which were measured in the same experimental condition. The LCM PL peak intensity at ~ 550 nm of the crystalline DA NPs increased approximately 4 times compared with the amorphous DA NPs. This phenomenon is originated from the crystal structure of the crystalline DA NPs [17].

## IV. CONCLUSION

Single crystalline DA NPs were fabricated through a solvent-vapor annealing method using the amorphous DA NPs which were fabricated through reprecipitation method. As observed in color CCD images and LCM PL spectra, the crystalline DA NPs exhibited bright and sharp light emission compared to the amorphous DA NPs. In the corresponding LCM PL spectra, the FWHM of the crystalline DA NPs was significantly narrowed and blue-shifted. These phenomena were originated from enhanced molecular interaction of the DA NPs in the single crystalline formation and effect of their nano-size.

**ACKNOWLEDGEMENTS**

This research was supported by National Research Foundation of Korea (NRF) grant funded by the Ministry of Education, Science and Technology (2013R1A1A1012445), the Inha University Research Grant (Grant No. INHA-51761), and Hwaam foundation. This work was supported by the National Research Foundation of Korea (NRF) grant funded by the Korean government (MSIP: Ministry of Science, ICT and Future Planning) (NRF-2015M2B2A8A07033186)

Figure Captions.

Figure 1. (a, b) HR-TEM images of the amorphous DA NPs and the crystalline DA NPs, respectively. (c, d) Magnification of HR-TEM images of the single unit of amorphous DA NP and the single unit of crystalline DA NP, respectively.

Figure 2. Comparison of WAXS patterns of the amorphous and crystalline DA NPs.

Figure 3. Color CCD images of the (a) amorphous and (b) crystalline DA NPs.

Figure 4. (a) Normalized LCM PL spectra ($\lambda_{ex}$ = 488 nm) of the amorphous and crystalline DA NPs. (b) Comparison of the LCM PL spectra intensity of the amorphous and crystalline DA NPs.

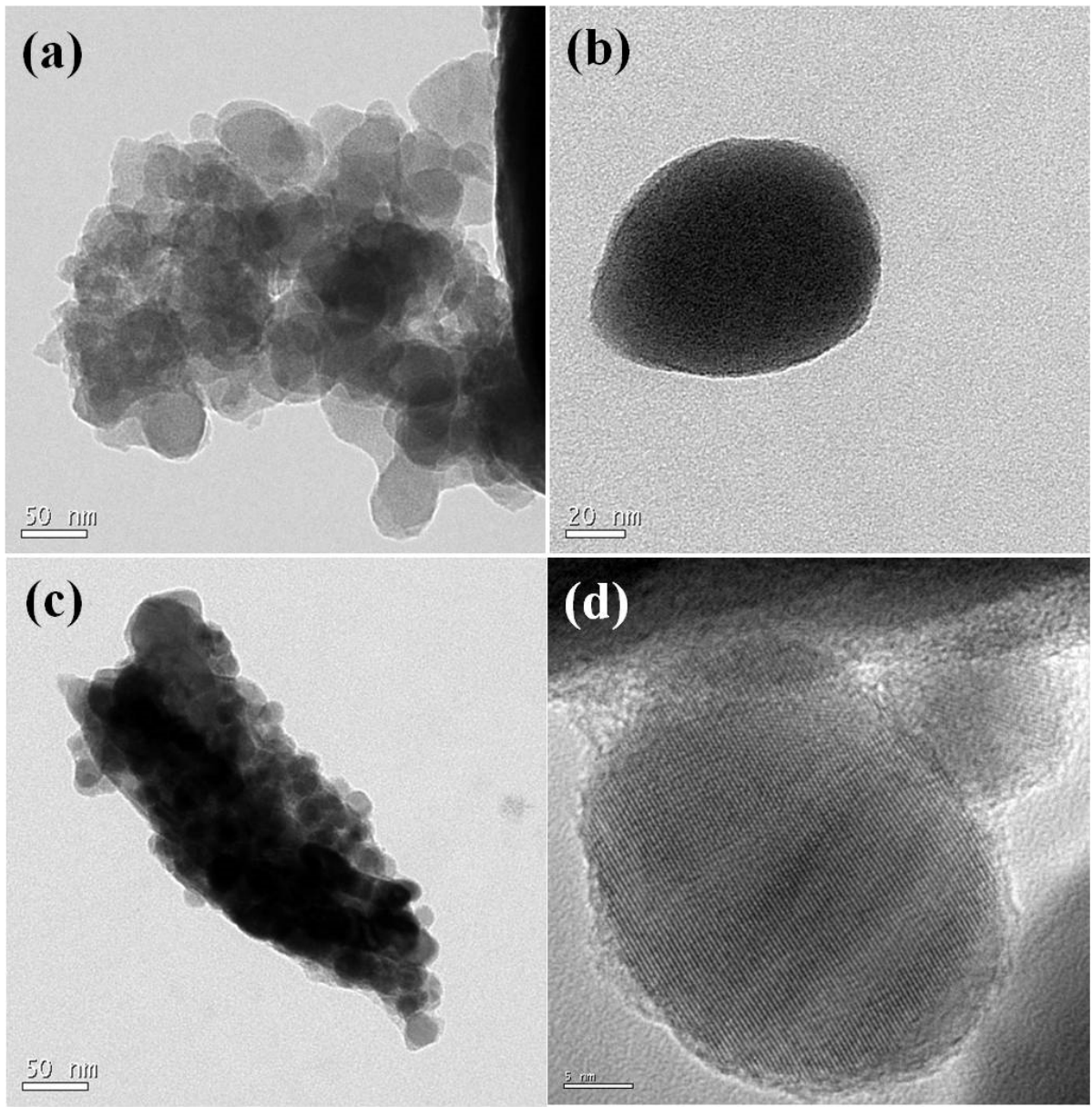

Figure 1 S. Kim *et al.*

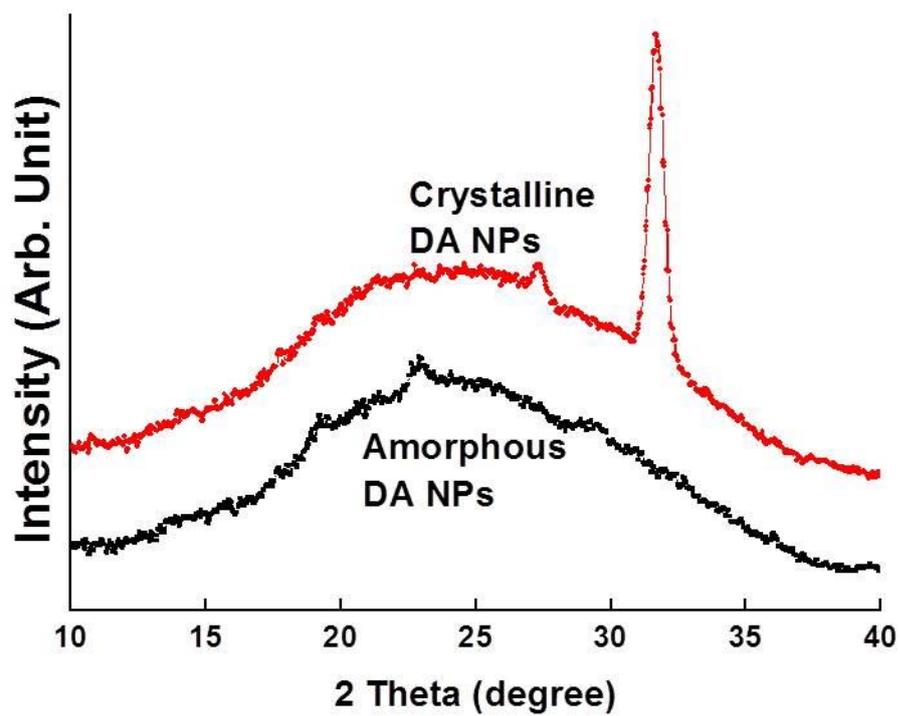

**Figure 2 S. Kim *et al.***

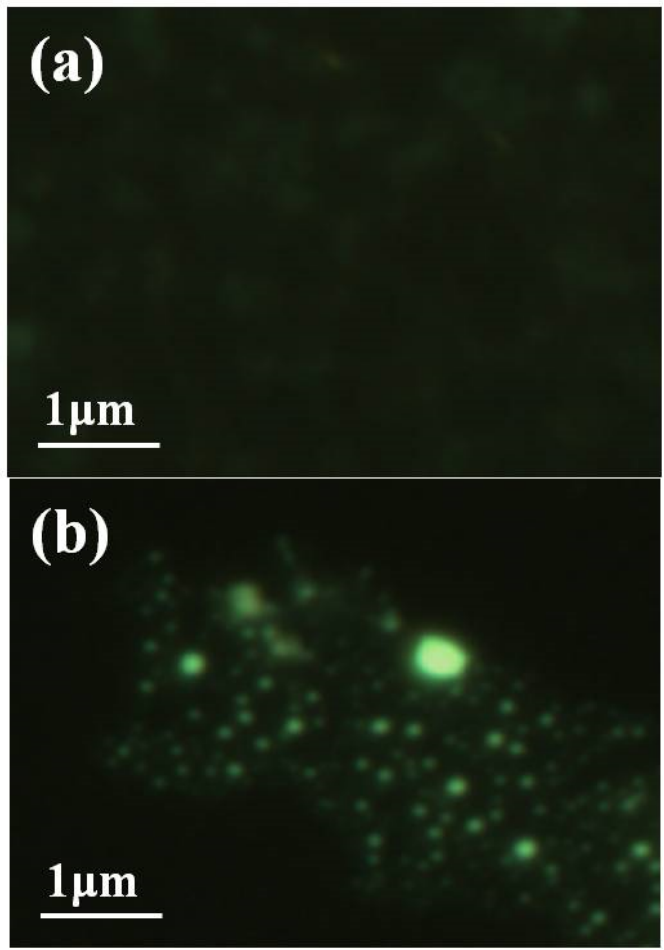

**Figure 3 S. Kim** *et al.*

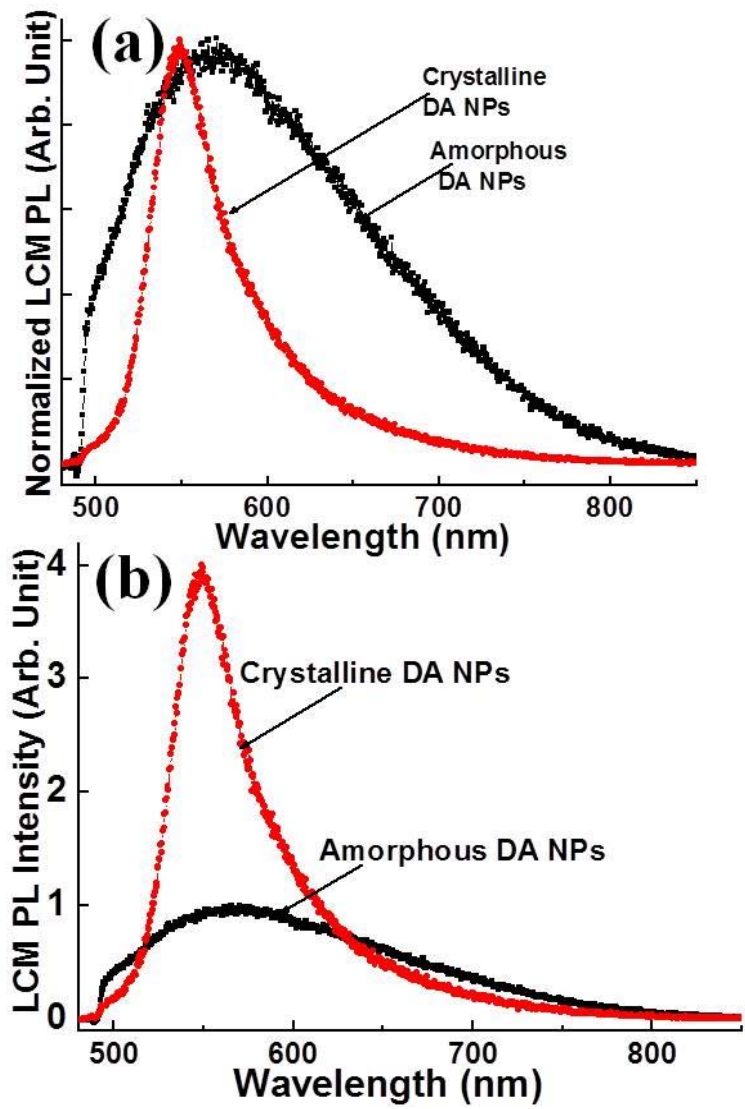

Figure 4 S. Kim *et al.*